\documentclass[aps,prx,twocolumn,floatfix,showpacs,a4paper]{revtex4-1}
\usepackage{graphicx}
\usepackage{amssymb}
\usepackage{amsmath}
\usepackage{bookmark}
\usepackage{hyperref}

\renewcommand{\phi}{\varphi}

\begin{document}

\title{Turing complete mechanical processor via automated nonlinear system design}
\author{Marc Serra-Garcia}
\email{sermarc@ethz.ch}
\affiliation{Institut f\"ur Theoretische Physik, ETH Zurich, CH-8093
Zurich, Switzerland}
\date{\today}

\begin{abstract}
Nanomechanical computers promise a greatly improved energetic efficiency compared to their electrical counterparts. However, progress towards this goal is hindered by a lack of modular components, such as logic gates or transistors, and systematic design strategies. This article describes a universal logic gate implemented as a nonlinear mass-spring-damper model, followed by an automated method to translate computations, expressed as source code of arbitrary complexity, into combinations of this basic building block. The proposed approach is validated numerically in two steps: First, a set of discrete models are generated from code. The models implement computations with increasing complexity, starting by a simple adder and ending in a 8-bit Turing complete mechanical processor. Then, the models are forward integrated to demonstrate their computing performance. The processor is validated by executing the Erathostenes' sieve algorithm to mechanically compute prime numbers.
\end{abstract}

\maketitle

\section{Introduction}

Nanomechanical computers \cite{sklan2015splash} have the potential for ultra-low energy information processing, which makes them attractive for implantable, wearable, remote or embedded applications where computational performance requirements are light but energy is scarce. They are also ideal for deeply cryogenic environments: In contrast with conventional electronics, nanomechanical resonators are unaffected by carrier freezeout phenomena, and dissipate extremely small amounts of heat (on the order of $pW$  \cite{barois2013ultra}, compared to tens of $\mu W$ for a high-mobility electrical transistor \cite{hemd_uw}), fitting the requirements for cryogenic quantum computing applications (e.g. as readout circuitry for spin qubits). While this potential has been known for quite some time \cite{roukes2004mechanical, Masmanidis780,merkle2018mechanical}, most recent works exploring mechanical information processing are limited to trivial computations, require complex geometries or heterogeneous materials, and do not provide a straightforward path to scalability \cite{liang2014acoustic, li2014granular, bilal2017bistable, malishava2015all,merkle2018mechanical,song2019additively}. This limitation exists despite the significant progress in systematic design of mechanical systems with thousands of degrees of freedom, driven primarily by the metamaterials community. In these recent works, the desired performance is first expressed as a set of symmetries \cite{Mousavi2015,PhysRevApplied.10.014017}, a stiffness or deformation map \cite{schumacher2015microstructures,RYS201931, Coulais2016} or a discrete mass-spring model \cite{MatlackPerturb,SerraGarciaQuadrupole}. Then, it is translated by a systematic algorithm into a device geometry that can be fabricated. However, this workflow is not sufficient for the goal of mechanical computation, because encoding a computation into symmetries, stiffnesses or discrete models is a profoundly hard problem, and explicit attempts to compute with metamaterials have been so far limited to linear operations such as differentiation or integration \cite{silva2014performing, estakhri2019inverse}.

\begin{figure}[!ht]
\centering
\includegraphics{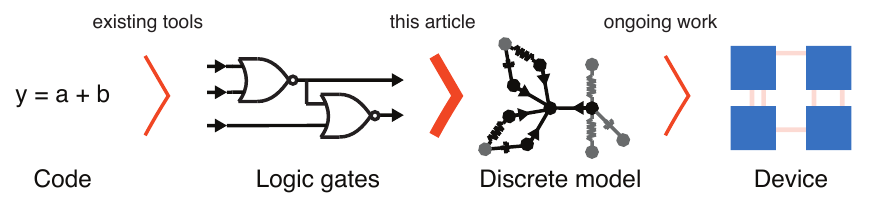}
\caption{Building mechanical logic from code. This paper addresses the problem of generating discrete models from a set of logic gates. Translating the resulting models into devices will not be discussed here, but has been investigated previously \cite{MatlackPerturb,SerraGarciaQuadrupole}, although future work will be required to address aspects such as nonlinearity and damping.}
\label{fig:fromcode}
\end{figure}

This paper addresses the problem of designing discrete models capable of performing complex computations. The proposed solution (Fig. \ref{fig:fromcode}) starts by writing the desired computation as code in a high-level language. This is much simpler than directly designing the discrete model, because code is a natural way to represent computations. Then, the code is translated into nonlinear mass-spring-damper models. This translation is performed in two steps: First, an existing, open-source tool \cite{yosys} is used to map the computation into a graph of elementary logic operations. Then, the elementary logic operations are replaced by mass-spring-damper systems with suitable input-output characteristics. The resulting models can have more than $10^4$ oscillating degrees of freedom, are not subject to any periodicity assumptions, and therefore are prohibitively complex to design without automation. 

In this work, computations will take place in the digital domain using a binary representation. Each bit of information will be encoded as the amplitude of vibration of a mass-spring resonator, with some range of amplitudes corresponding to a binary zero and some range of amplitudes corresponding to a binary one.  It should be noted that there is no requirement that a binary zero be represented by a near-zero amplitude. Through this work, $x$ and $y$ will denote logical variables,  which can only take 0 or 1 values. Physical variables will be denoted by $u$ and $F$, corresponding to displacements and forces respectively. Physical variables can take a continuum of values and will generally follow harmonic trajectories. Inputs will be implemented by applying a harmonic force, with angular frequency $\omega$, to an input degree of freedom, $F(t) = F_{0} sin(\omega t)$, where $F_{0}$ will be chosen from the zero or one ranges to implement a zero or one input respectively. Outputs will be determined by monitoring the vibration amplitude of an output degree of freedom, and checking whether it lies in the "zero" or "one" range. Here, the range for zero is defined as being between $36.5\%$ and $67\%$ of a reference force $F_R$ or displacement $u_R$ (See appendix), while one is defined as being between $92\%$ and $102.5\%$ of the reference. This also provides a recipe to compose logical functions: they should be connected with a spring of strength $k_C=F_R/u_R$, so an output vibration amplitude in the zero or one range will produce an input force in the zero or one range respectively.

The paper will be structured in three parts. The first part will focus on combinatorial logic circuits, whose outputs are a function of the current inputs only, and do not depend on the past input-output history. Combinatorial circuits will be built by combining instances of a basic building block using automated design tools. The section will cover the design process of the building block, including the requirements that it must satisfy in order to be able to combine a large number of them to form an advanced logical function. In the second part, sequential circuits will be considered, whose output depends on both the current input and previous input history. For these, an additional element incorporating memory will be introduced. In the third part, a full processor will be demonstrated. In contrast with the previous examples, where a discrete model can solve a single problem, the processor can be programmed to solve different problems by setting the initial conditions.

\section{combinatorial logic}

\begin{figure}[!ht]
\centering
\includegraphics{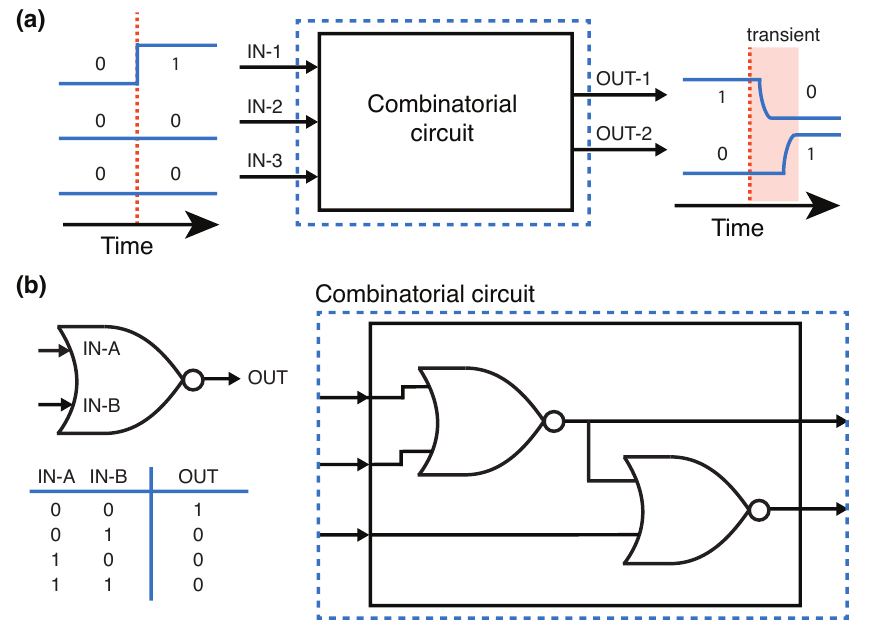}
\caption{Combinatorial logic circuits a) Input-output response of a combinatorial system with three inputs and two outputs. When the inputs are changed, the outputs eventually converge to the correct value, but may take incorrect values during the transient period. These temporary, incorrect values are called \emph{hazards } in the language of digital electronics. b) NOR gate schematic symbol and input-output response (truth table) [left] and potential realization of the combinatorial circuit in (a) by combination of NOR gates [right].}
\label{fig:combi_intro}
\end{figure}

Combinatorial logic circuits are those whose steady-state output $\vec{y}(t\rightarrow\infty)$ is an arbitrary logical function of the input vector $\vec{x}$, $\vec{y}(t\rightarrow\infty)=f(\vec{x})$. When the input is changed to a new state $\vec{x}_1$, the output will eventually converge to the new updated $f(\vec{x}_1)$. In the interval right after the input is changed, there will be a transient and the output may temporarily take incorrect values (Fig. \ref{fig:combi_intro}a). 

It is a well-known result that any arbitrary logical function $f(\vec{x})$ can be implemented as a network of realizations of a single universal logic gate (Fig. \ref{fig:combi_intro}b). Therefore, building a combinatorial mechanical computer requires, minimally, finding a mechanical analog of such universal gate (plus potentially some rules on how gates should be ``wired'' together). The resulting implementation is not unique, and the network providing the best speed of computation may not be the one that requires the smallest number of logic gates. In this work, all systems will be built by combining instances of a mechanical NOR gate. Having access to a broader set of building blocks may result in increased performance, and software mapping logical functions into graphs of gates will attempt to use all available gate types to reach pre-defined goals of complexity and speed. 

\subsection{Requirements for the basic building block}

Prior works in mechanical logic have focused on simple systems consisting of isolated logic gates \cite{Masmanidis780, liang2014acoustic} or, at most, pairs of them \cite{bilal2017bistable}. This masks emergent phenomena that arise in the presence of a large number of gates. Since the objective of this work is to create arbitrarily complex computations, it becomes crucial to design a building block that is robust when combining a large number of instances. This modularity can be formalized into a set of six requirements, similar to those considered by the photonic computing community \cite{miller2010optical}:

\begin{enumerate}
\item  A free-standing basic building block must not present multistability under valid logical inputs, that is, the steady-state output must be known from its inputs, without regard to the past input-output history. The requirement follows directly from the definition of combinatorial logic, however, it is not met by some of the proposals in the literature \cite{bilal2017bistable}.  This requirement does not apply when the inputs are outside the valid zero or one range or if the building block is embedded in a network containing logic loops, with the later configuration being used to implement memories in the \emph{sequential logic} section of this paper.
\item The basic building block must operate at a single frequency, meaning that the frequency of the output must be the same as the frequencies driving the inputs. This is necessary to allow concatenation of multiple logic operations. Such requirement was not met by some earlier proposals for mechanical logic \cite{li2014granular}, but has since then been recognized \cite{bilal2017bistable}.
\item The building block must be capable of \emph{fanout}, i.e., driving multiple inputs from a single output. This requirement is well-understood in photonic computing \cite{miller2010optical}, but has not been explicitly demonstrated in mechanical logic.
\item  The inputs must be ``easy to drive'', meaning that an input's force-displacement response at the operating frequency must be a simple (ideally linear) function and not have complex dependencies on the other inputs and the output. The motivation behind this requirement is to enable the computationally-efficient combination of a large number of building blocks. When this requirement is met, gates in a logical network can be optimized individually, by lumping all other elements into a single mass-spring-dashpot. The difficulty in meeting this requirement is one of the reasons why automated design of mechanical systems has been thought to be extremely hard or impossible \cite{whitney1996mechanical}.
\item The block must produce \emph{digital level reconstruction} \cite{miller2010optical}, meaning that (sufficiently small) deviations from the exact zero and one values should be self-correcting, resulting in an output that is closer to the correct value than the inputs.
\item The building block must be insensitive to the phase of the oscillation driving its inputs, because such dependence imposes the onerous requirement than total phase shifts be controlled to less than a quarter of a period over the whole length of a computation. Phase-sensitive interference effects were originally proposed for mechanical logic \cite{Masmanidis780} but have since been understood by the photonics community to be critically difficult to realize \cite{miller2010optical}.
\end{enumerate}

In addition, and while this work is exclusively numerical, some additional requirements will be considered to ensure that the resulting discrete models are not too dissimilar from what can be accessed in existing experimental platforms: 

\begin{enumerate}
\item The building block must be constructed using a single type of nonlinearity, which must be low-order. This ensures that the resulting equations are not stiff, constrains the search space, and makes the system more universal as low-order nonlinearities appear in a broad range of experimental platforms, offering multiple avenues for experimental realization.
\item The block must contain a small number of degrees of freedom, in order to minimize simulation or fabrication costs. This is in contrast with some works in the field of mechanical logic \cite{li2014granular, bilal2017bistable}, that attempt to mimic conventional transistors based on bulk lattice phenomena, and therefore intentionally utilize a large number of degrees of freedom.
\end{enumerate}

A potential ninth requirement, namely that the performance of the system be insensitive to small changes in the properties of the building block, will not be considered here, but may be of prime importance for experimental realizations where fabrication defects are unavoidable.

\subsection{NOR gate design and principle of operation}

\begin{figure}[!ht]
\centering
\includegraphics{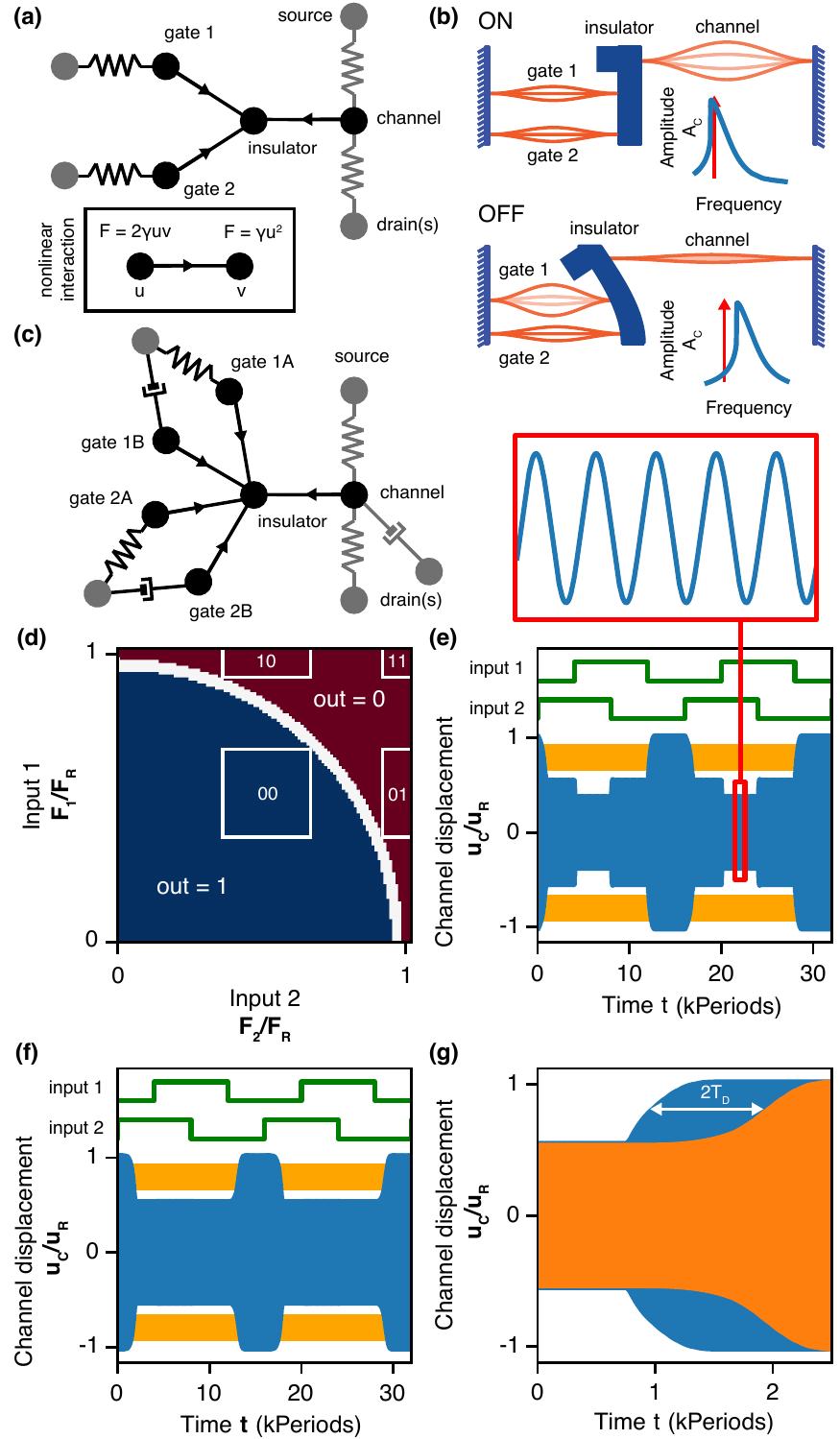}
\caption{Building block for mechanical logic. a) Simplified discrete model for a mechanical NOR gate. Circles represent harmonic oscillators, connected by linear and nonlinear (inset) springs. Grayed-out degrees of freedom are shared between building blocks (i.e the drain of a building block will actually be the gate of a different block). b) Implementation of (a) with strings (orange) and a cantilever (blue), and channel resonance (inset) with excitation amplitude (red arrow).  c) Full discrete model for the NOR gate. d) Steady-state relation between the channel vibration amplitude and the forces acting on the input gates. The color corresponds to the channel vibration ampltiude (red = zero region, blue = one region, white = ambiguous region). e) Channel oscillation under varying gate excitation amplitudes. The shaded area denotes an invalid (ambiguous) output. f) Output (channel) displacement for a system of three cascaded NOR gates. Deviations from the correct amplitude levels are reduced after each NOR application. g) Transient response of the one-gate (blue) and three gate (orange) system, the delay is $T_D=500 periods/gate$. }
\label{fig:combi_buildingblock}
\end{figure}

The basic building block or logic gate (Shown in a simplified form in Fig. \ref{fig:combi_buildingblock}a, b and in full form in Fig. \ref{fig:combi_buildingblock}c) acts as a ``valve'', allowing or blocking the flow of mechanical energy between a \emph{source} and a \emph{drain} harmonic oscillators, depending on the presence of an excitation on its input terminals, and according to the gate's truth table. Here we refer to the input terminals as \emph{gates}, in analogy with field-effect transistors, whose control terminal is called the gate. The distinction between gate as input terminal and gate as basic logic building block should be obvious from context. 
The source and drain terminals are not directly connected by springs. Instead, they are both connected to a \emph{channel} harmonic oscillator. When the channel's natural frequency is in resonance with the excitation frequency, the channel's vibration amplitude is high and this results in a large energy transfer between source and drain. When the channel is off-resonance, there is little motion and lower source-drain energy transfer. The logic gate operating mechanism consists in using nonlinearity to shift the natural frequency of the channel depending on the presence or absence of excitation at the gates.

Gate-dependent adjustment of the channel's resonance frequency is accomplished by means of a nonlinear interaction that couples the gates and channel through an intermediate harmonic oscillator called the insulator (Fig. \ref{fig:combi_buildingblock}a). The insulator degree of freedom prevents the direct flow of mechanical energy between inputs and outputs, allowing only for a gate-mediated channel frequency modulation that takes place on a slower time scale than the gate/channel oscillation. The gate-insulator and insulator-channel nonlinear interaction is derived from a potential of the form $H=\gamma u^2 v$ (See inset in Fig. \ref{fig:combi_buildingblock}a). This type of low-order nonlinear interaction is very common in nature, arising in optomechanical systems \cite{aspelmeyer2014cavity}, magnet lattices \cite{serra2018tunable} and vibrating strings \cite{serra2016mechanical} among others. 

To provide an intuitive understanding of the principle of operation behind the building block, it is worth looking at an implementation based on vibrating strings and cantilevers (Fig. \ref{fig:combi_buildingblock}b), similar to the setups in refs \cite{goto1959parametron, serra2016mechanical}. It should be noted that this implementation is provided to aid in the understanding and may not be an optimal avenue to the experimental realization of mechanical computers. In this implementation, the gates and the channel are vibrating strings. The insulator is a cantilever to which gates and channel are attached.   When the gates are driven by a large harmonic force, their amplitude of oscillation increases. As a consequence, their tension increases (vibrating strings spend more time in a curved configuration that is longer, and therefore has higher tension, than an equivalent string at rest). The gates' dynamic tension bends the insulator (cantilever), and adds tension to the channel, increasing its resonance frequency, as in the tuning of a guitar string. This tuning/detuning mechanism drives the channel in and out of resonance, and is responsible for the gate-mediated modulation of the source-drain energy transfer.

The equations of motion for a free-standing building block are given by:

\begin{align}
\mathcal{H}_C-2\gamma_{IC}u_{I}u_C&= F_{C} sin (\omega t) \label{eq:eqmotion_simple}\\
\mathcal{H}_{I}-\gamma_{IC}u_C^2+\gamma_{IG}u_{G1}^2+\gamma_{IG}u_{G2}^2 &=0\\
\mathcal{H}_{G1}+2\gamma_{IG}u_{I}u_{G1}&= F_{G1} sin (\omega t)\\
\mathcal{H}_{G2}+2\gamma_{IG}u_{I}u_{G2}&= F_{G2} sin (\omega t)
\end{align}where the subindices $C$, $I$, $G1$ and $G2$ denote the channel, insulator and the two gates respectively, the excitation frequency is $\omega$, and the nonlinear couplings between insulator-channel and insulator-gate are given by $\gamma_{IC}$ and  $\gamma_{IG}$. $\mathcal{H}_{X}=0$ is the equation of motion for a simple harmonic oscillator, i.e.:

\begin{eqnarray} \label{eq:eqmotion_harm}
\mathcal{H}_X=m_X\ddot{u}_X+\frac{m_X\omega_{X}}{Q_X}\dot{u}_X+m_X\omega_{X}^{2}u_X,
\end{eqnarray}where, for a degree of freedom $X \in \{C, I, G1, G2\}$, $m_X$ is the oscillating mass, $Q_X$ is the quality factor and $\omega_{X}$ is the natural frequency. All parameters are provided as an appendix, as well as in the supplementary code files.

The system does not include equations of motion for the source and drain. This is because in the digital designs considered here, these will be typically shared among various devices. The source of all devices will be a single high-mass degree of freedom used to input energy into the system, and will oscillate at constant amplitude. Therefore, once coupled to the channel, it will act as a harmonic force given by $F_{C} sin (\omega t)$, with $F_C = k_{SC}A_S$ with $k_{SC}$ being the source-channel spring constant and $A_S$ being the source amplitude. When concatenating multiple devices, the role of the drain will be played by the gate of a building block downstream.

\subsection{Numerical simulation of the NOR gate}

Equations \ref{eq:eqmotion_simple}-\ref{eq:eqmotion_harm} are integrated using a 4th order Runge-Kutta algorithm written in C++ \cite{press2007numerical}, with a time step of $0.03-0.025$ periods of oscillation, after verifying that such time step is sufficiently small to converge to the correct solution within plotting precision. The C++ file containing the equations of motion is generated and compiled from a high-level description of the system's degrees of freedom and interactions, written in Python (\emph{Jupyter}). The steady-state response can also be calculated semi-analytically, by prescribing the location of the insulator degree of freedom, calculating the amplitudes of the gates and channel by treating them as linear harmonic oscillators, using the calculated amplitudes to determine the mean force acting on the insulator, and iterating until static equilibrium is attained, i.e. $\langle F_I \rangle =-\gamma_{IC} \langle u_C^2 \rangle +\gamma_{IG} \langle u_{G1}^2+u_{G2}^2 \rangle =-K_Iu_I$. All code files, which enable the rapid exploration of mechanical logic devices, are provided as a supplementary material. 

Numerical simulations of an isolated building block, as presented in Fig. \ref{fig:combi_buildingblock}a, b and described by equations \ref{eq:eqmotion_simple}-\ref{eq:eqmotion_harm}, produce satisfactory results (not shown) in terms of input-output characteristics, digital level reconstruction and lack of multistability. However, attempts at connecting multiple building blocks to implement advanced logic functions have been unsuccessful. The cause of this failure is that the basic building block, as described in Fig. \ref{fig:combi_buildingblock}a, b violates the requirement of being "easy to drive", i.e. the gate stiffness is highly dependent on the insulator displacement due to back-action of the gate-insulator nonlinear coupling. As a consequence, when connecting two building blocks (using a linear spring to connect the channel of the first block to the gate of the second block), the effective stiffness of the channel in the first block becomes sensitive to the second block insulator location. In these conditions, it has not been possible to find a value for the system's parameters that achieves the right operating conditions for all insulator locations. This issue highlights a core motivation behind the present paper: Combining a large number of building blocks gives rise to unexpected issues that may not be easy to foresee, and therefore proposals for mechanical logic should be tested in sufficiently complex examples.

\subsection{Scalable coupling of multiple building blocks}

A solution to the stiffness back-action issue is presented in Fig. \ref{fig:combi_buildingblock}c. It consists in using two physical gates for every logical input. When cascading multiple blocks, the two gates are connected asymmetrically: One with a linear spring, presenting a force-displacement relation $F=k_C(u_2-u_1)$, and one with a dashpot, described by a force-velocity relation $F=(k_C/\omega)(\dot{u}_2-\dot{u}_1)$. (Here $k_C$ is the coupling strength, and $u_1$ and $u_2$ are the two coupled degrees of freedom, typically a channel and a gate). Since in a harmonic oscillator the velocity is shifted by $\pi/2$ with respect to the displacement, the force acting on the dashpot-driven gate will be shifted by $\pi/2$ with respect to the spring-driven gate, resulting in a phase shift of $\pi/2$ between gates. For the dashpot gate, the corresponding back-action force done by the gate into the channel will accumulate an additional shift of $\pi/2$ for the same reason, giving a total phase shift of $\pi$. In contrast, the back-action force from the spring-coupled gate will not suffer either phase shift. Therefore, gate back-actions will differ by a factor of $\pi$ and experience destructive interference. By adjusting the spring and dashpot coupling constants to produce forces of identical magnitude, it is possible to create an aggregate back-action that, at the frequency of operation, is insensitive to the insulator location. 

From an energy conservation point of view, this destructive interference must be understood as the system consuming a maximal amount of energy at all times, irrespectively of whether this energy is dissipated at the gates or at the coupling damper. Experimentally, physical dashpots are not necessary to produce destructive interference. The effect can also be obtained by using waveguides with lengths differing by $\lambda/4$. In addition, it may be possible to obtain a more robust cancellation by using ideas from topological physics \cite{huber2016topological}, where there are known examples of mass-spring systems that are immune to stiffness changes acting symmetrically on pairs of resonators \cite{SusstrunkPendulaScience}. However, these approaches are not investigated here, because they use a much higher number of degrees of freedom.

The building block's operation as NOR gate is validated by numerical simulations of its steady-state (Fig. \ref{fig:combi_buildingblock}d) and transient (Fig. \ref{fig:combi_buildingblock}e) responses. The steady-state response in Fig. \ref{fig:combi_buildingblock}d demonstrates the digital level reconstruction that is desired of a digital component. Even significant deviations from an acceptable input will result in an output that lies in the correct range, though the margin is thin for a small range of amplitude pairs near the transition region. These results have been obtained by performing only cursory optimization of the system's parameters, suggesting the potential for greater margins and a much more robust noise immunity.  The tendency of the system to reconstruct deviations from ideal logic values can be observed by concatenating multiple logic gates. Figure \ref{fig:combi_buildingblock}f depicts the output of the logical function $y(x_1, x_2) = (((x_1\:\text{NOR}\:x_2)\:\text{NOR}\:0)\:\text{NOR}\:0)$, which is equivalent to $y(x_1, x_2) = x_1\:\text{NOR}\:x_2$ depicted in Fig. \ref{fig:combi_buildingblock}f, but where the signal has gone through two additional gates. It can be seen that the output presents much less variability than the result of a single gate.

The ability to produce a correct digital output for a broad range of input values has its origins in the steep, asymmetric frequency-amplitude response of the channel, which can be seen in the inset of Fig. \ref{fig:combi_buildingblock}b. This asymmetry arises from a feedback mechanism that can be understood easily in the string-cantilever representation of  Fig. \ref{fig:combi_buildingblock}b: When the channel's amplitude decreases, its own dynamic tension decreases. As a consequence, the cantilever experiences lower bending resistance and displaces even more, increasing the resonance frequency shift and causing additional amplitude decrease. Using this nonlinear mechanism for digital level reconstruction requires paying attention to several aspects: First, the location of the steep jump is highly sensitive to damping. This is not an issue in numerical simulations, where damping can be prescribed exactly, but may be a problem for experiments where damping is typically hard to control. Second, the mechanism can introduce bistability, as it occurs in optomechanical systems \cite{aspelmeyer2014cavity}. Here it has been numerically observed that bistability does not occur when the gate excitation amplitude lies in the ranges corresponding to a valid 0 or 1. Third, near the bifurcation point where the system transitions between having a single and multiple solutions, some of the effective time constants governing convergence to the steady-state become increasingly long (being infinite at the exact bifurcation point, where the system has no tendency to converge between the two stable solutions). While exploring this phenomenon is outside the scope of this work, a discussion on this issue for a related system can be found in the supplementary material of Reference \cite{serra2016extreme}. Here, it is confirmed numerically that the speed of convergence is not significantly slowed down by nonlinearity as the system is essentially fully converged after $3Q_C$ periods (Fig. \ref{fig:combi_buildingblock}g). While these three issues do not pose significant challenges in a numerical study, future experimental realizations may benefit from considering alternative mechanisms of response steepening for digital level reconstruction, for example the use of composite channels containing more than one harmonic oscillator.  

\subsection{Automatic synthesis of combinatorial circuits}

Once it has been established that the basic building block meets the requirements for digital computation, it is possible to implement advanced logic functions in an automated manner. In this paper, this is done in the following approach: First, the target computation is described using the high-level language Verilog. Then, the open-source tool Yosys \cite{yosys} is used to map the high-level description into a graph of logic gates. The graph of logic gates is then translated into a graph of harmonic oscillators, connected via linear springs, nonlinear springs and dashpots, by replacing each logic gate by the corresponding mass-spring model as depicted in Fig. \ref{fig:combi_buildingblock}c. The coupling strength for the inter-gate springs and dashpots is determined using the relation between reference force and reference amplitude introduced previously. Finally, the resulting dynamical system is forward-integrated numerically to confirm its ability to perform correct computations.

\begin{figure}[!ht]
\centering
\includegraphics{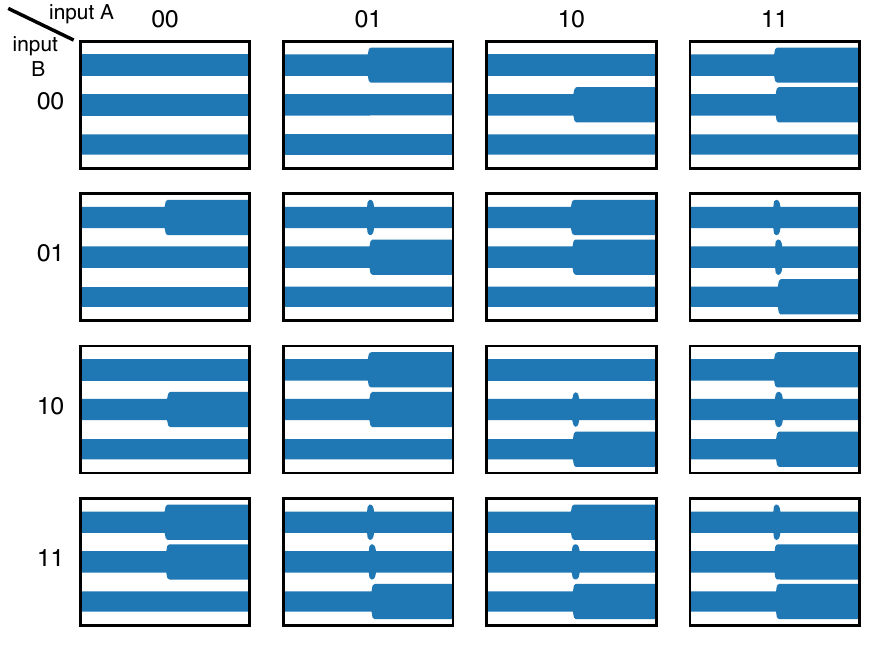}
\caption{Numerical simulation of a mechanical adder. Each panel plots the displacement of the three outputs (with the trace corresponding to the least significant bit on the top). The inputs of the adder are zero during the first half of the simulation, and then are suddenly set to the value given by the row and column.  Dynamic hazards can be observed in some of the transitions.}
\label{fig:combi_synthesis}
\end{figure}

This approach is first demonstrated for a digital adder that takes two 2-bit numbers and generates a 3-bit result containing the sum of the two operands. The Verilog code for the adder consists of three lines of code, and results in 17 logic gates that translate to 102 harmonic oscillators interacting via 19 linear springs, 85 nonlinear springs and 19 dashpots. Figure \ref{fig:combi_synthesis} shows a numerical simulation of the adder for all possible inputs.

\section{Sequential logic}

Combinatorial logic allows the realization of mechanical systems implementing arbitrarily complex functions. However, it presents several limitations: First, it does not offer a notion of memory or state, which is crucial in many computing applications. Second, the number of gates required to realize a multi-step computation may be prohibitively high, as the implementation must have independent, separate logical circuits for every step of the computation. Sequential logic, where the output of a system will depend on the previous history in addition to the inputs, presents a solution to these problems. The dependence on previous history will introduce a means to store information, and it will be possible to re-use the same logical circuit for multiple steps of the same computation, by performing the steps at different points in time and using the previous state dependence to store intermediate results. There are several approaches to sequential logic. This paper will focus on synchronous sequential logic, where state transitions are driven by a periodic clock. 

\begin{figure}[!ht]
\centering
\includegraphics{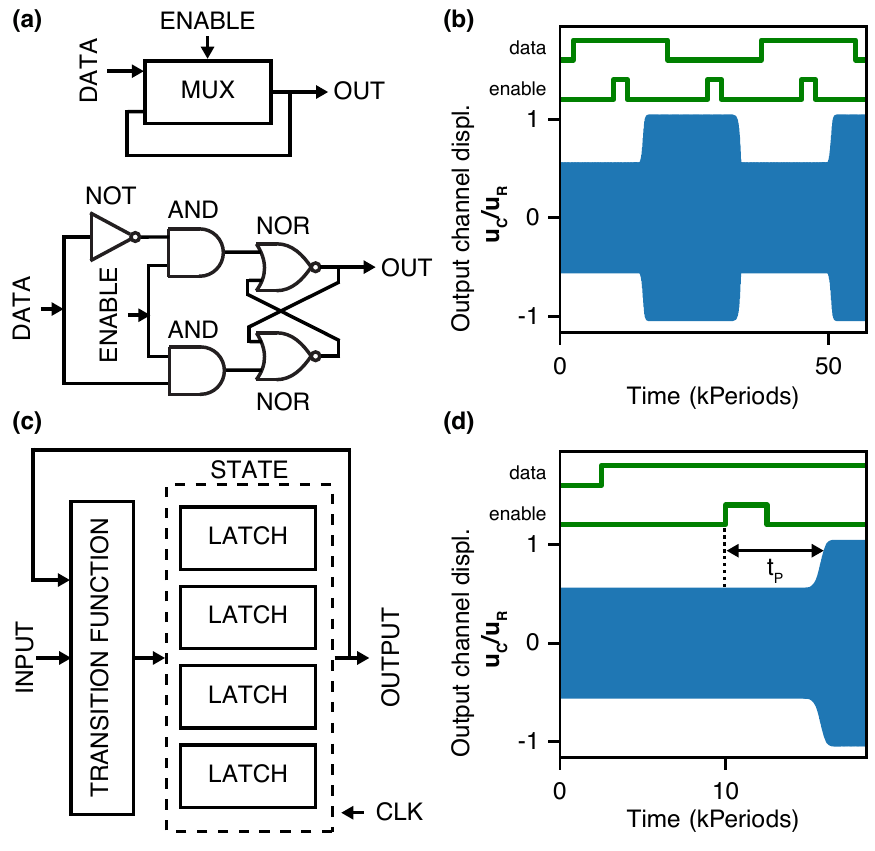}
\caption{Mechanical implementation of a digital latch. a) The latch can be understood as a multiplexer (MUX) that selects between keeping the same output or updating the output with the contents of the \emph{data} input (top). Implementations are highly symmetrical to avoid output hazards that might prevent latching (bottom). b) Numerical simulation of the mechanical latch, implemented in terms of the building block introduced in Fig. \ref{fig:combi_buildingblock}. When the enable input is pulsed high, the output is updated to match the data input. c) Finite state machine, implemented by combining a combinatorial transition function and a set of latches to store the state. d) Propagation delay in the latch, ensuring that the output does not update until the latching pulse has ended. }
\label{fig:sequential_dlatch}
\end{figure}

\subsection{Latches and flip-flops}

Implementing sequential logic requires the introduction of a new building block capable of storing information. This device will have two inputs and an output. One of the inputs will contain the data to be stored, and the second input will indicate when the data should be committed to memory. The output will contain the current data in the block, and will change upon receipt of a \emph{store} command. Unless such command is received, the output will be insensitive to changes in the data input. In conventional digital electronics, two classes of devices are commonly employed for this task: Latches and flip-flops. Latches store information whenever the \emph{enable} input is at digital level 1, while flip-flops store information when the \emph{clock} input switches from 0 to 1.  Flip-flops are more common than latches in conventional electronics, but here we will use latches as basic building blocks because they are simpler.

A latch can be implemented using a two-way multiplexer (MUX, Fig. \ref{fig:sequential_dlatch}a). A two-way multiplexer is a combinatorial device that has one output and three inputs (\emph{A}, \emph{B}, and \emph{select}). When \emph{select} is low, the output will be equal to \emph{A}, while when \emph{select} is high, the output will be equal to \emph{B}. The logical function that describes the multiplexer is $f(A,B,select)=(A \;\text{AND}\;(\text{NOT}\:select))\;\text{OR}\;(B\;\text{AND}\;select)$. The latch is created by introducing a loop between the output and input \emph{A}. Now, when \emph{select} is low, the output will stay at the same value. When \emph{select} is high, the output will change to match the input. The latch implemented in this work follows a somewhat more symmetric design \cite{ndjountche2016digital} (Fig. \ref{fig:sequential_dlatch}a), that prevents issues due to output hazards. The mechanical AND and NOT gates in Fig. \ref{fig:sequential_dlatch}a are implemented by combining instances of the NOR gate defined in Fig. \ref{fig:combi_buildingblock}c. Numerical simulations of a dynamical system implementing a latch are presented in Fig. \ref{fig:sequential_dlatch}b.

\subsection{Finite state machines}

One of the most common applications of sequential logic is in the realization of \emph{finite state machines}. Finite state machines are characterized by being in one state out of a finite set, and transitioning to a different state in response to an external event or after an interval of time, according to its transition function. An example finite state machine is shown in Fig. \ref{fig:sequential_dlatch}c. In this example, the state is stored in a set of 4 latches. A combinatorial transition function determines the next state, as a function of the current state and external inputs into the machine. When the clock 'ticks', presenting a value of 1 at the enable inputs of the latches, the output of the combinatorial function gets stored and becomes the current state. For finite state machines to operate correctly, it is necessary that the output of the transition function remains constant during the latching interval. This is accomplished by using a relatively short latch pulse of 2500 periods (Fig. \ref{fig:sequential_dlatch}d) and adding an additional delay to the output of the latch. Because of this intentional delay, the output of the latches (and therefore the inputs and outputs of the transition function) will not start changing until the latching pulse has already ended (Fig. \ref{fig:sequential_dlatch}d). The delay is implemented by concatenating 5 NOR gate pairs, each pair performing an identity function, $I(x) = (x\:\text{NOR}\:0)\:\text{NOR}\:0$, but adding a delay of its own. The total delay is measured to be on the order of 6500 periods. This is consistent with the $T_D=500\;periods/gate$ determined in Fig.  \ref{fig:combi_buildingblock}g, as signal must propagate through an AND gate (composed of 2 cascaded NOR gates), a NOR gate, plus 5 gate pairs for the output delay, adding to a total of 13 NOR gates.

\begin{figure}[!ht]
\centering
\includegraphics{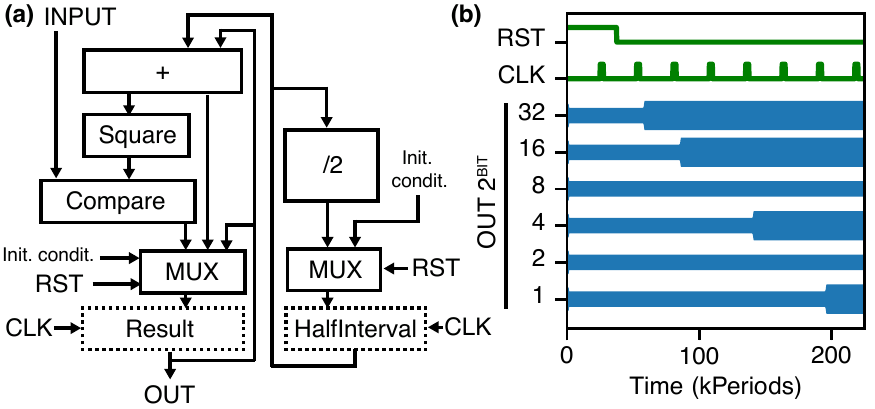}
\caption{Calculation of the square root with a mechanical finite state machine a) Architecture of the machine, based on binary search. The solid rectangles denote combinatorial functions, while the dotted ones represent registers (groups of latches that store a state). The current guess for the square root (initialized at zero) is combinatorially increased by half the search interval, and then is combinatorially squared.  A multiplexer chooses the next guess between the initial condition of zero (when reset equals one), the current guess (if adding half the interval causes the square to exceed the input value) or the current guess plus half the interval (if the calculated square is equal or below the target value). The search interval is halved at every step, and a multiplexer chooses between the halved interval (under normal operation) or the maximum interval, to prepare for a new computation (when reset equals one). When the clock ticks, the next result and next interval are latched and become the current state. b) Numerical simulation of the machine during calculation of the square root of 2809. At every clock cycle, one bit of the result converges to the final value, until the correct value of 53 is reached.  }
\label{fig:sequential_squareroot}
\end{figure}

\subsection{Automatic synthesis of sequential circuits}

The Verilog-Yosys workflow that was used to generate combinatorial logic can be extended to the sequential case. This is accomplished by adding a flip-flop element to the Yosys library and assigning it to the \emph{latch} device of Fig. \ref{fig:sequential_dlatch}a. The latch will store data every time the clock signal pulses, in contrast to a real flip-flop, that would store it when the clock signal presents an edge. However, this does not have an impact in the performance of the designs, provided that the clock pulse is short enough and that only positive or negative edges are used as triggers in the Verilog code. This limitation can be overcome by utilizing two separate clock signals to emulate positive or negative edge triggers, or by implementing a genuine flip-flop using loops of combinatorial elements. To illustrate the capability of automatically synthesizing sequential logic in a nontrivial computation, the next example will demonstrate a mass-spring-dashpot system that calculates square roots. The example operates by binary search. It starts by partitioning the search interval in half. The value at the middle of the search interval is then combinatorially squared and compared to the number whose square root is being calculated. Depending on the result of the comparison, the search interval is restricted to the upper half or the lower half of the previous search interval. Figure \ref{fig:sequential_squareroot}a shows the architecture of the square root calculator. The example, written in Verilog, is 15 lines of code long and generates a nonlinear mechanical system containing 3006 harmonic oscillators interacting via 685 linear springs, 2505 nonlinear springs and 685 dashpots. The system is numerically integrated and the results are shown in Fig. \ref{fig:sequential_squareroot}b. Latches are driven by pulses that are 2500 periods long, with 25000 periods of pause between pulses to ensure enough time for the signal to propagate through the combinatorial network. After $2*10^5$ periods of oscillation, the result has converged to the correct value.

\section{Demonstration of a mechanical processor}

The finite state machine introduced in Fig. \ref{fig:sequential_squareroot} can solve a single problem, that is, calculating square roots. However, it is possible to create universal processors that can evaluate arbitrary functions. These contain logical circuits capable of performing a range of operations, called \emph{instructions}. The processor determines which instruction is executed by reading a \emph{stored program} from a memory, and can also use the memory to hold intermediate results. This section will cover a mass-spring-dashpot model describing a mechanical processor. 

\begin{figure}[!ht]
\centering
\includegraphics{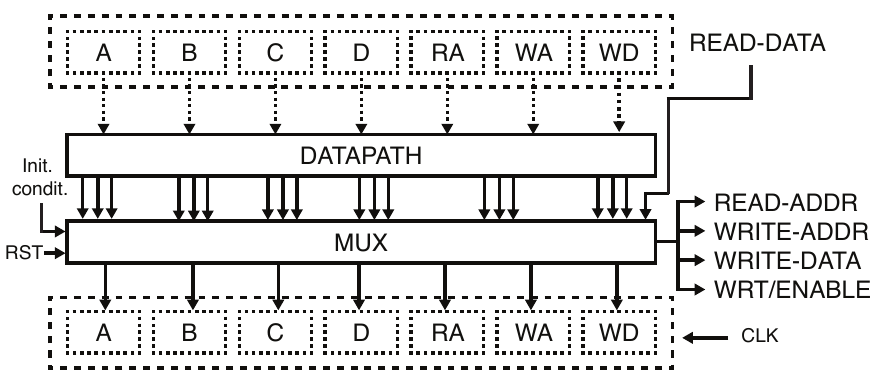}
\caption{Architecture of the mechanical processor. The processor contains a set of registers implemented with latches (dashed rectangles). A combinatorial function (datapath) determines all the potential future values for the registers. A multiplexer (MUX) selects which of the values should be stored in the registers, depending on the instruction currently being read from memory. When the clock "ticks", the future value of the registers gets latched and becomes the present value.}
\label{fig:pt}
\label{fig:sequential_procarc}
\end{figure}

The processor cannot solve problems stand-alone, it also requires a memory where the program, intermediate values and final results are stored. Processor and memory can be thought of as black-boxes communicating through a set of shared degrees of freedom. These degrees of freedom are split into two separate ports: A read port, that the processor uses to ask the memory for information, and a write port, that the processor uses to store information in the memory. The processor's read port consists of 8 output degrees of freedom (containing a number indicating which of the 256 memory addresses is being read) and 8 input degrees of freedom (that receive an oscillatory force from from the memory, encoding the data stored at the requested memory address). The write port contains 17 output degrees of freedom. Eight of them contain the data to be written, another 8 contain address where the data should be written, and the last one indicates when the output data is valid and should be committed to memory.  Mechanically, all outputs will be channel oscillators from a building block like the one in Fig. \ref{fig:combi_buildingblock}c, whose amplitude of vibration will communicate the address and data to the memory. The inputs will be gate oscillators, and be driven by a harmonic force from the memory. Here, only the processor will be simulated numerically. The memory will be emulated by adding an additional function to the numerical simulation that actively monitors the output degrees of freedom, and generates excitation forces according to the requested address and the contents of a data array. There are no fundamental restrictions preventing the realization of a mechanical memory composed of latches, but this is not done here because it results in a very high number of degrees of freedom ($> 10^5$) and makes numerical simulation extremely time-consuming. 

While open-source processor designs written in Verilog exist, here a custom design is introduced. This is done for two reasons: First, since the processor has to be numerically simulated in full, it is crucial to have an absolutely minimum number of logic gates. This is typically not as serious of a consideration in electronics, as physical transistors are inexpensive. Second, most electronic designs make use of a feature called \emph{tri-stating}, where the same line can act both as an output (being set to states '0' or '1') or as an input (by being set to what's typically called state 'Z'). While there is no fundamental reason for tri-stating to be impossible in a mechanical system, that would require an additional building block whose development is outside the scope of this paper. The processor introduced here has 6 internal 8-bits registers, each of them implemented as a set of 8 latches as described in Fig. \ref{fig:sequential_dlatch}. The first three registers will contain the memory address currently being read, the memory address currently being written, and the data currently being written. The remaining four registers, labelled A, B, C, D will contain the operands and results of the instruction currently being executed. The architecture of the processor is as follows: A combinatorial \emph{datapath} section calculates the potential next values of the seven memories for all potential instructions. Then, a multiplexer chooses which of the potential values should be committed to the register, based on the instruction currently present at the data input (which is read from the main memory). Figure \ref{fig:sequential_procarc} ilustrates this architecture.

The processor can execute the following 16 instructions divided into four categories. Register management instructions are: \emph{SWAP{\_}AC}, \emph{SWAP{\_}BD}, \emph{SWAP{\_}AB}: Swap the contents of the A-C, B-D and A-B registers respectively,  \emph{COPY{\_}AC}: Copies the content of A into C, leaving A intact. Load/store instructions: \emph{SAVE{\_}AB}: Stores the content of register A in the memory address contained in register B. \emph{LOAD{\_}AB}: Loads the content of the memory address specified in register A into register B.  \emph{LDNX{\_}AB}: Loads the contents of the memory address right after that of the instruction currently being executed, and places them in register A. Instruction execution will continue at the current address + 2. Flow control instructions: \emph{SKPNX{\_}A{\_}GRT{\_}B}: Skips the next instruction if the value in register A is greater than that in register B, \emph{SKPNX{\_}A{\_}DIF{\_}B}: Skips the next instruction if the content of register A are different from that of register B, \emph{SKPNX{\_}OVER}: Skips the next instruction if the last arithmetic addition resulted in an overflow, i.e. the sum of the two operands exceeded 255. \emph{JMP{\_}C}: Resumes the execution at the memory address given by the register C. Arithmetic and logic instructions: \emph{NOT{\_}C}: Replaces the content of register C by its bit-wise negation, \emph{ANORB{\_}TO{\_}C}: Computes a bit-wise NOR between registers A and B and stores it in register C, \emph{APLUSB{\_}TO{\_}D}: Computes the sum of register A and B and stores the result in register D. \emph{RSL{\_}D}: Computes a logical right shift of register D (all bits are shifted to the right, and the most significant bit is replaced by zero). \emph{RSA{\_}D}: Computes an arithmetic right shift of register D: All bits are shifted to the right, but the most significant bit is left intact. 

The mechanical processor spans 128 lines of Verilog code. It synthesizes to 11754 harmonic oscillators interacting via 2744 linear springs, 9795 nonlinear springs and 2744 dashpots. 

\begin{figure}[t]
\centering
\includegraphics{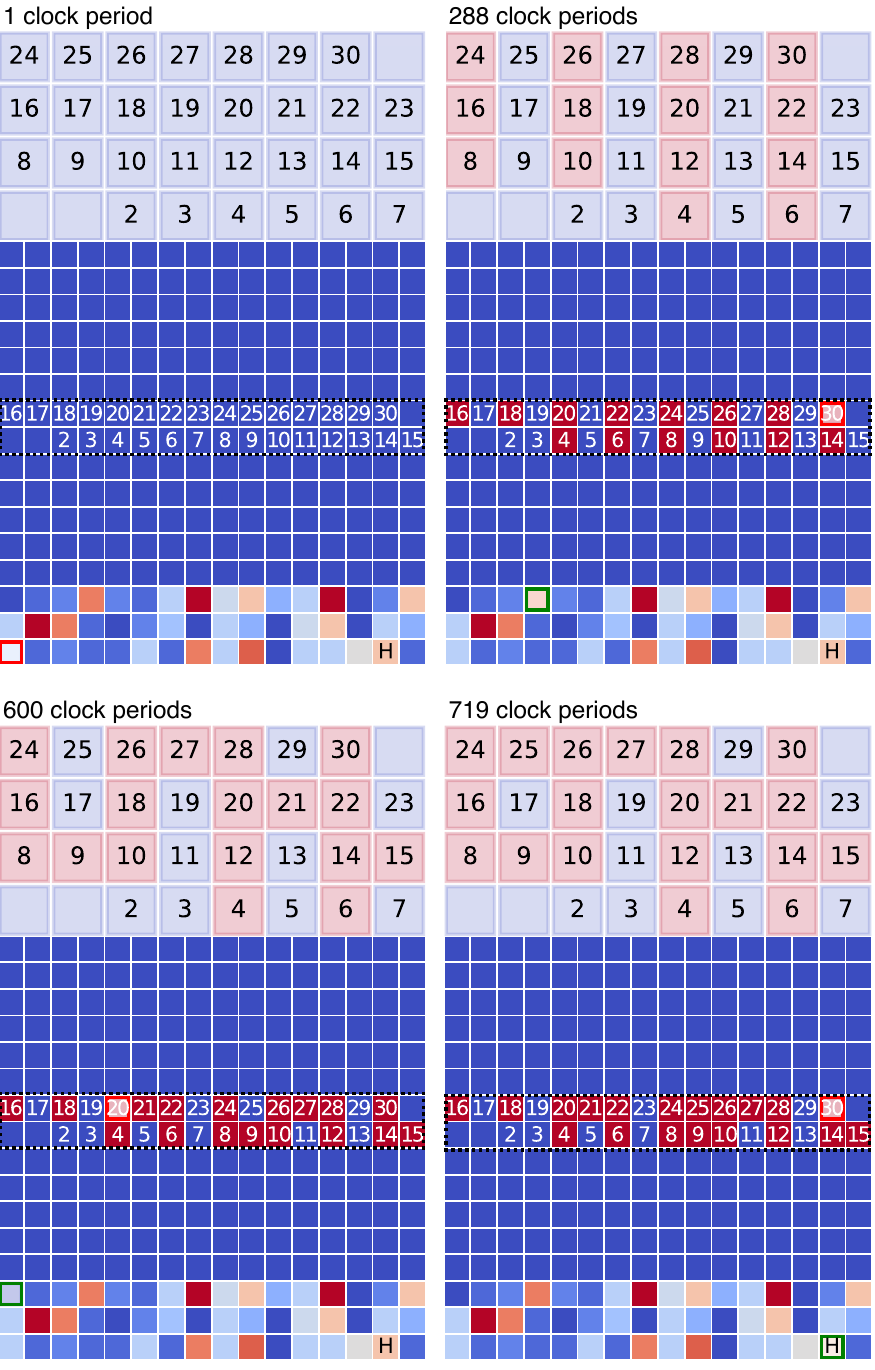}
\caption{Numerical simulation of a mechanical processor during the execution of Erathostenes' Sieve algorithm. Each of the four panels depict the contents of the system's memory at a different point during the execution of the algorithm. The 16x16 small squares represent each memory address, where the color indicates the value stored. The green rectangle indicates the memory address currently being read, and the red rectangle indicates the memory address currently being written. Memory addresses grow first along rows and then through columns, with the zero address corresponding to the bottom left corner. The bottom three rows are the 48 bytes that contain the instructions of the program. The region starting at address 128 (dotted rectangle) is where the primality of numbers is stored. Its value changes during the execution. The top 32 larger squares in each panel present a magnified view of this memory region.  }
\label{fig:sequential_erathostenes}
\end{figure}

The mechanical processor is tested by executing the Erathostenes sieve algorithm to produce prime numbers smaller than $N_{\text{MAX}}$ (here taken to be 32). This particular problem is chosen because prime numbers rarely appear in dynamical system (in contrast with e.g. additions or sine waves). Therefore, the ability to mechanically generate them demonstrates the flexibility of the approach. The algorithm works as follows: An array (Here represented by memory addresses 128 - 160) will store the primality of numbers. The array will be initialized at zero, indicating that all numbers are presumed prime to begin with. The algorithm will start by checking the smallest prime number, 2. It will keep adding the number to itself and marking the resulting memory address as composite, until the end of the array is reached. Then, the next number will be considered. If the number hasn't been found composite, the same marking process will take place, if the number has been found composite, the marking process will be skipped and the next number will be considered. Only numbers up to $\sqrt{N_{\text{MAX}}}$ need to be considered. The algorithm is 48 bytes long, including instructions and constants (e.g. memory addresses, initial values and ranges). The code of the algorithm, together with an execution trace, is provided as a supplementary information. The code for the algorithm is loaded into the processor by setting it as initial condition for the emulated memory. Then, the processor's dynamics are numerically simulated for 719 clock periods, until the algorithm reaches the halt condition. The simulation takes 50h 22 minutes to complete on a Core i7 laptop from 2015, indicating that, while a complex task, these problems can be handled without access to supercomputing resources. Figure \ref{fig:sequential_erathostenes} shows the contents of the system's memory at different points during the execution of the algorithm.

The Erathostenes algorithm example ilustrates how an engineered mechanical system can produce an output, prime numbers, that does not naturally appear in dynamical systems. However, the question remains on how general is the class of problems that this processor can solve. The answer is that any \emph{computable} function can be evaluated with such a device. This is captured by the notion of Turing-completeness:  Once a processor attains a certain level of complexity, adding additional instructions may make it more efficient, but does not increase the number of problems that it can solve. For example, multiplication can be simulated by repeated addition, and therefore a processor does not need dedicated multiplying logic to be able to solve problems involving products. The same is true for exponentiation, which can be emulated by repeated multiplication. In order to evaluate arbitrary \emph{computable} functions though, a processor needs access to an unbounded memory. Since unbounded memory does not exist in practice, the label Turing-complete is generally used to refer to processors that could evaluate any computable functions if they were to be augmented by an unbounded memory. Turing completeness is proven in Appendix B by constructing an emulator for a known \emph{Universal Turing Machine}.

\section{Conclusions}

The numerical results in this paper demonstrate nonlinear mass-spring-damper models capable of performing complex computations, ranging from a simple two-bit adder to a Turing-complete processor. This has been accomplished by designing a highly modular building-block implementing a basic logic operation, and utilizing existing tools to map advanced computations into instances of this basic logic operation. While this work demonstrates a route towards advanced mechanical information processing, two crucial obstacles must be overcome for its experimental realization: First, there is no automated mechanism to generate geometries implementing the resulting discrete models. The design approach introduced in the context of perturbative metamaterials \cite{MatlackPerturb,SerraGarciaQuadrupole} provides a route towards this goal, but is so far limited to linear systems. Second, the systems discussed here are highly dependent on hard-to-control parameters such as damping. More robust building blocks need to be found, which will probably require the use of additional degrees of freedom. These two problems cannot be considered separately, the optimization of the building block must be done once an experimental platform has been established, and a realistic range of parameters and uncertainties has been determined.

The set of examples developed here are sufficiently complex to demonstrate the flexibility of the approach, but sufficiently simple to be simulated numerically without overwhelming computational requirements, and can be used as a benchmark for future works involving mechanical logic. The importance of performing these tests has also been highlighted by the observation that apparently-functional building blocks, such as the one presented in Fig. \ref{fig:combi_buildingblock}a, cannot be scaled into more complex designs due effects that are hard to foresee. Finally, here, a set of mechanical systems have been generated automatically from a code description of their intended behavior. This approach is extremely successful in the field of electrical engineering, where integrated circuits containing billions of transistors are generated from source code descriptions of their functionality. However, the possibility of applying a similar design methodology to mechanical systems has been the subject of a long-standing debate\cite{whitney1996mechanical, antonsson1997potential}. The results presented here support the position that, in the light of recent advances in mechanical modelling and optimization\cite{Mousavi2015,PhysRevApplied.10.014017, schumacher2015microstructures,RYS201931, Coulais2016, MatlackPerturb,SerraGarciaQuadrupole}, the automated design of ultra-complex mechanical systems can soon become a reality.

\section{Acknowledgements}

The author would like to thank Tena Dub\v{c}ek, Sebastian Huber, Pascal Engeler and Eli\v{s}ka Greplova for helpful comments on the presentation of the manuscript.

This work has received funding from the European Research Council under the Grant Agreement No. 771503, from the Swiss National Science Foundation and from the NCCR QSIT.

\bibliographystyle{phd-url}
\bibliography{ref}

\section{APPENDIX A: System parameters}
The gate parameters are $m_G=0.2[M]$, $Q_G=200$, $k_G=m_G\omega_G^2=25.582[F]/[L]$. Insulator parameters are  $m_I=1.0[M]$, $Q_I=1.5$, $k_I=m_I\omega_I^2=0.394784[F]/[L]$. Channel parameters are: $m_C=6.0[M]$, $Q_C=1.5$ $k_C=m_C\omega_C^2= 14311.8[F]/[L]$. Nonlinear couplings are $\gamma_{IG}=12[F]/[L^2]$ and $\gamma_{IC}=6[F]/[L^2]$ The reference forces and displacements are $F_R=0.7[F]$ and $u_R=0.0182[L]$. The channel excitation is $F_C=1.326[F]$. Exceeding these force values may result in diverging simulations due to the gate or insulator stiffness becoming negative. Local parameters must stay the same when cascading devices to form logical networks, so the local stiffness and damping have to be adjusted to compensate for the contribution of the coupling springs and dashpots. This may result in negative values. While these are attainable experimentally (e.g. through magnetic forces, buckling or stored elastic energy for the stiffness and through parametric or optomechanical pumping for the damping), further optimization to remove them may be advisable before experimental realization. 

\section{APPENDIX B: Proof of Turing completeness}
Turing-completeness is proven if the processor is able to simulate a known \emph{Universal Turing Machine} (UTM), as UTMs are characterized by being able to evaluate any computable function. A UTM $U(m,n)$ is a finite-state machine that can be in one out of $m$ states. The machine has an unbounded tape where the program is stored. Each cell of the tape contains a symbol from a set of $n$ symbols. At every iteration, the machine reads a symbol from a position (typically called the head location), and, as a function of the current state and last read symbol, performs the following three tasks: Writes a new symbol in the current head location, moves the head to right or to left, and transitions to a new state. The particular Universal Turing machine considered here has $n=5$ possible symbols and $m=5$ possible states \cite{neary2009four}). 

To show that the processor presented here is Turing-complete, the processor must be augmented with some mechanism to access external storage. In digital electronics, a common way to accomplish this is to use memory-mapped registers (MMR). A MMR is a memory address that is singled out and used to provide input and output to the system. The processor does not distinguish between MMR and conventional addresses. However, when data is read from or written to the MMR, the data is sent to an output or read from an external input rather than from the main memory. To implement a MMR, a combinatorial multiplexer is inserted between the processor and the memory. When the read or write address is different from the MMR address, information is directed to and from the memory. However, if the read or write address matches the MMR address, read data is taken from the system's input, and write data is sent to the external output. A MMR can be used to access the tape of the universal Turing machine. The 8-bit output register can be used to control the tape, with three bits representing the symbol to be written (only 5 of the 8 possible values will be used) and one bit representing the direction of motion for the tape. Then, the current symbol under the head can be read using the input MMR. Once a means to control the external tape has been established, a universal Turing machine can be simulated in a straightforward manner: The memory is initialized so one 25-byte region, the ``output table'', contains a  byte $o_i$ indicating the symbol to be written and movement to be performed, for each of the 25  possible combinations of current state and read symbol. The output corresponding to the state $p_i$ and last-read symbol $s_j$ will be stored at the address $a = a_{out} + i*n+j$, where $i,j\in\{0, 1, 2, 3,4\}$. Another 25-byte region is initialized to contain the next state as a function of the current state and read symbol (here called the ``transition table''), at addresses $a = a_{trans} + i*n+j$. The state $p_i$ will be represented by $a_{out} + i*n$ to simplify emulation. Then, assuming that the initial state is contained in a processor register, the emulation algorithm is as follows: First, the current symbol is read by loading the MMR into another processor register. The two registers are added together to calculate a location in the "output table". The content of that location is read and saved into the memory-mapped register to produce the required output/motion. Then, the memory distance between the transition table and the output table $a_{trans}-a_{out}$ is added to the location in the output table, to generate the location in the transition table. The contents from the generated address are read, and become the new current state. At this moment, the algorithm jumps to the initial position.

\end{document}